\documentclass[fleqn,10pt,usenames,dvipsnames]{wlscirep}

\usepackage[utf8]{inputenc} 
\usepackage[T1]{fontenc}
\usepackage{amsmath,amssymb,amsfonts}
\usepackage{graphicx} 

\usepackage{amsthm}
\usepackage{cite}
\usepackage{xr-hyper} 
\usepackage{hyperref}
\usepackage{nameref}
\usepackage{url}
\usepackage{booktabs,tabularx,ltablex,longtable}
\usepackage{eucal} 
\usepackage{bbold} 

\usepackage{multirow}
\usepackage[oldenum]{paralist}
\usepackage{enumitem}
\usepackage[left]{lineno}

\newcommand{\avg}[1]{\left\langle #1 \right\rangle}

\newcommand{\modu}[1]{\bigl\lvert #1 \bigr\rvert}
\newcommand{\tempor}{\mathcal{T}}
\newcommand{\avgtempor}{\overline{\mathcal{T}}}
\newcommand{\avgtemportheo}{\overline{\mathcal{T}}_{\text{th}}}
\newcommand{\avgtemporrand}{\overline{\mathcal{T}}_{\text{rand}}}
\newcommand{\timewind}{\Delta \tau}
\newcommand{\timewindstar}{\Delta \tau^\star}
\newcommand{\baboonsdata}{\texttt{Baboons}}
\newcommand{\malawidata}{\texttt{Malawi}}
\newcommand{\highschooldata}{\texttt{High School}}
\newcommand{\sfhhconfdata}{\texttt{SFHH Conference}}
\newcommand{\hospitaldata}{\texttt{Hospital}}
\newcommand{\tradturcarpdata}{\texttt{Trade - carpets Turkey}}
\newcommand{\traditagunsdata}{\texttt{Trade - guns Italy}}
\newcommand{\traditaceradata}{\texttt{Trade - cereals Italy}}
\newcommand{\braindata}{\texttt{Brain}}
\newcommand{\emaildata}{\texttt{E-mails}}
\newcommand{\usflightsdata}{\texttt{US domestic flights}}

\newcommand{\ie}{\emph{i.e.},}
\newcommand{\eg}{\emph{e.g.},}
\newcommand{\etal}{\emph{et al.}}

\newcommand{\ktot}{K_{\text{\textsc{TOT}}}}
\newcommand{\commtrade}{\textsc{COMMTRADE}}

\DeclareMathAlphabet{\mathcal}{OMS}{cmsy}{m}{n}

\graphicspath{{figures/}}

\title{Characterization of interactions' persistence in time-varying networks}

\author[1,2]{Francisco Bauz\'a Mingueza}
\author[3,2]{Mario Flor\'ia}
\author[3,2]{Jes\'us G\'omez-Garde\~nes}
\author[4]{Alex Arenas}
\author[4,2,5,*]{Alessio Cardillo}

\affil[1]{Department of Theoretical Physics, University of Zaragoza, E-50006 Zaragoza, Spain}
\affil[2]{GOTHAM Lab -- Institute for Biocomputation and Physics of Complex Systems (BIFI), University of Zaragoza, E-50018 Zaragoza, Spain}
\affil[3]{Department of Condensed Matter Physics, University of Zaragoza, E-50006 Zaragoza, Spain}
\affil[4]{Department of Computer Science and Mathematics, University Rovira i Virgili, E-43007 Tarragona, Spain}
\affil[5]{Internet Interdisciplinary Institute (IN3), Open University of Catalonia, E-08018 Barcelona, Spain} 
\affil[*]{acardillo@uoc.edu}

\begin{abstract}
Many complex networked systems exhibit volatile dynamic interactions among their vertices, whose order and persistence reverberate on the outcome of dynamical processes taking place on them. To quantify and characterize the similarity of the snapshots of a time-varying network -- a proxy for the persistence,--  we present a study on the persistence of the interactions based on a descriptor named \emph{temporality}. We use the average value of the temporality, $\overline{\mathcal{T}}$, to assess how ``\emph{special}'' is a given time-varying network within the configuration space of ordered sequences of snapshots. We analyse the temporality of several empirical networks and find that empirical sequences are much more similar than their randomized counterparts. We study also the effects on $\overline{\mathcal{T}}$ induced by the (time) resolution at which interactions take place.
\end{abstract}

\begin{document}


\flushbottom
\maketitle

\thispagestyle{empty}

\section*{Introduction}
\label{sec:intro}

Over the last decades, complex networks have been used successfully to study a wide range of complex systems: from biological to technological systems, from social to economical ones just to cite a few~\cite{boccaletti-phys_rep-2006,barabasi-nphys-2012}. Nevertheless, the evolving nature of many complex systems at timescales of specific studies still requires of quantification tools~\cite{book-masuda-2020,zhang-epj_b-2017,holme-phys_rep-2012}.

The evolution of interactions between agents in complex systems over time does not only affect the structural properties of networked systems~\cite{Granell15}, but also the dynamics taking place on them. Indeed, it has been found that time-varying interactions change the behaviour of dynamical processes like: epidemic spreading~\cite{gross-prl-2006,masuda-prl-2013,liu-prl-2014}, diffusion~\cite{perra-prl-2012}, synchronization~\cite{lucas-epl-2018,kohar-pre-2014,frasca-prl-2008}, pattern formation~\cite{petit-prl-2017}, and evolutionary game theory~\cite{cardillo-pre-2014}. In particular, the speed of the variation of the interactions plays a pivotal role on the outcome of the dynamics~\cite{masuda-scirep-2016}. In some cases, the time-scales at which the interactions and the dynamics evolve are distinct~\cite{darst16}, allowing the system to be studied under either the \emph{quenched} (\ie{} static) approximation~\cite{meyers-jtheobio-2005,kao-roy_soc_int-2007} or the annealed one~\cite{boguna-pre-2009,guerra-pre-2010}. The former approach is more suitable when the dynamics evolves much faster than the network's structure (which can be thought as if it is static), whereas the latter approach (which leverages a well-mixing approach) is more appropriate in the opposite scenario, since it is equivalent to the case where every individual contacts a sufficient number of individuals to have information on the overall system state. More often, the two time-scales are not distinguishable~\cite{moody-soc_for-2002,Bagrow-pone-2011,pu-science-2009}, thus requiring more sophisticated techniques to study first, and understand then, the phenomenology observed.

A key feature of the evolution of interactions is their degree of \emph{persistence} (particularly at a short-range in time). It is known that the persistence has effects on different types of dynamics such as evolutionary game theory~\cite{cardillo-pre-2012,li-natcomms-2020}, synchronization~\cite{fujiwara-pre-2011}, and diffusive processes~\cite{starnini-pre-2012,masuda-prl-2013}, as well as on the properties of communication patterns among people~\cite{godoy_lorite-pone-2016}. Beyond that, recent studies have highlighted a clear relationship between the interactions' persistence and the differentiation, in networks, between the main backbone of the interactions and the noise -- or spurious interactions, -- that simply ``switch on and off''~\cite{kobayashi-natcomm-2019,presigny-pre-2021}.

Thus, the intricate interplay between the evolution of the interactions and the dynamics taking place on a complex network calls for a deeper understanding of time-varying interactions' characteristics, with special attention on their persistence. One way to achieve such a goal would by answering to the question: \emph{How special/rare is the observed temporal order of the interactions occurring in a time-varying network}? To solve this conundrum, we propose to set:
\begin{inparaenum}[i)]
\item A null hypothesis/model to be used for generating a benchmark interactions' order.
\item A meaningful property capturing the features of the temporal order of the interactions of a time-varying network.
\end{inparaenum}

\begin{figure}[t!]
\centering
\includegraphics[angle=-90,width=0.9\textwidth]{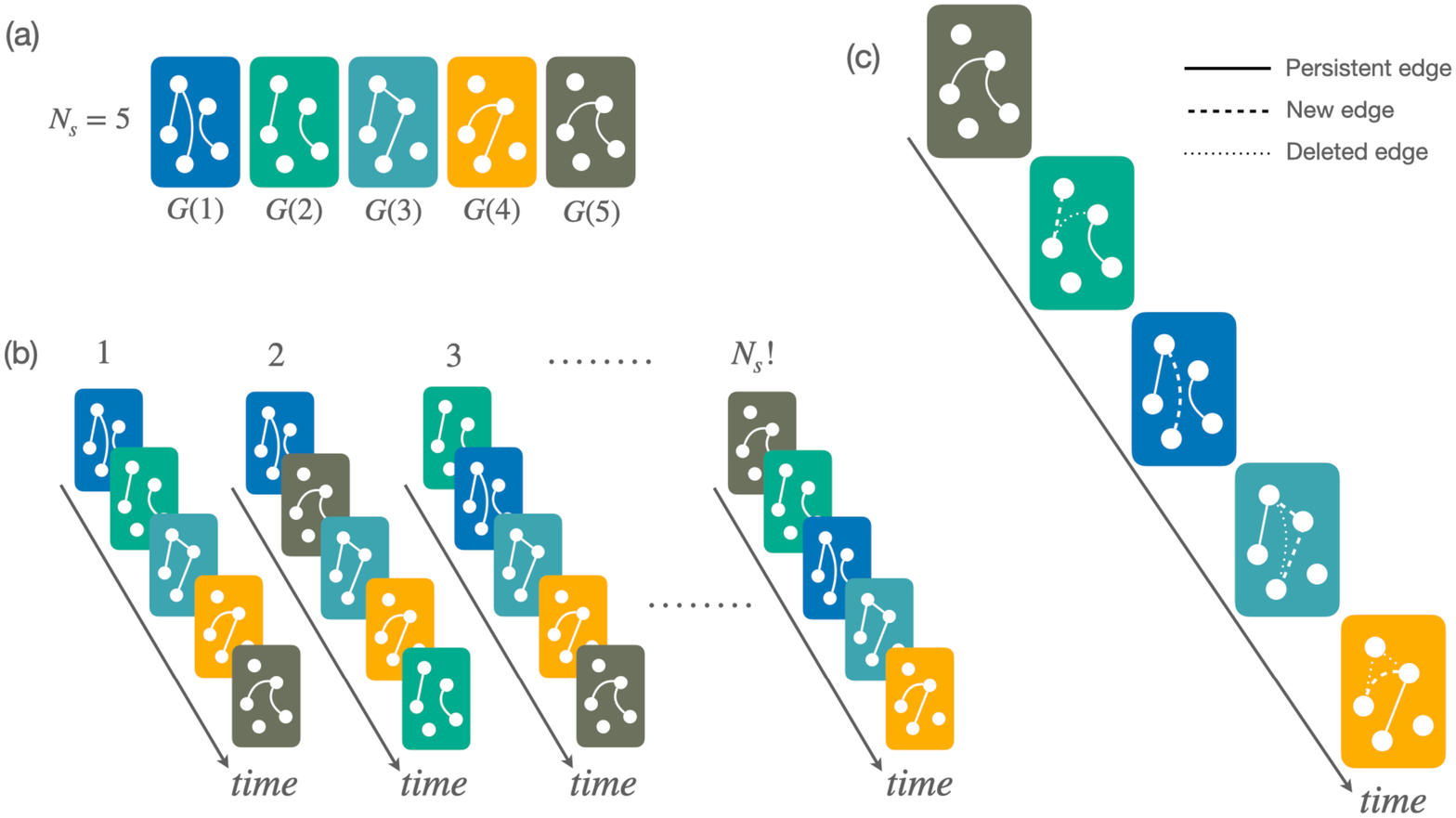}
\caption{Schematic illustration of the basic features of our framework. (a) A simple collection of $N_s = 5$ snapshot graphs constituting the building-blocks of a time-varying network. In (b), the former snapshots are randomly shuffled yielding different instances of the time-varying graphs. The real graph corresponds to one of the $N_s!$ possible orderings. Finally, in (c) we analyse a given sequence of snapshots comparing each snapshot with the previous one, highlighting: (i) edges that persist (solid lines), (ii) the new edges (dashed lines), and (iii) removed edges (dotted lines).}
\label{fig:tgv_scheme}
\end{figure}

Concerning the former, we assume that the underlying and invariant feature of a time-varying network is the set of interactions taking place at a given instant \ie{} the set of $N_s$ snapshot graphs (see Fig.~\ref{fig:tgv_scheme}.a) constituting the time-varying network. The ordered sequence of snapshots is a well-defined ordination out of all the possible ones. Such a relationship can be thought as the outcome of a shuffling process taking place in the configuration space of all the possible ordinations, with a specific sequence corresponding to one point of such a space~\cite{masuda-scirep-2016,masuda-prl-2013}. The volume of the configuration space is equal to the number of possible configurations $N_{\text{conf}}$, which -- in the absence of additional assumptions, -- is the number of permutations of the $N_s$ snapshots, $N_s!$ (see Fig.~\ref{fig:tgv_scheme}.b). Under these premises, assessing how special/rare a time-varying network is, corresponds to nothing else than computing the probability of obtaining an ordination with a given property within the configuration space. This physical-statistical framework -- and the idea of a random reshuffling of time snapshots as a generator of the configuration space, -- takes advantage of a group of models and processes, known within the literature on the topic with the name of Microcanonical Randomized Reference Model (MRRM) (see~\cite{gauvin-siam-2022} and references therein).

Regarding the choice of a descriptor encapsulating the features of the snapshots' ordination, an ideal candidate should be able to capture one of the main properties of the latter: how interactions -- \ie{} the set of edges, -- evolve from one snapshot to the next one (see Fig.~\ref{fig:tgv_scheme}.c). Over the years several metrics have been proposed (see \cite{masuda-scirep-2019,zhan-arxiv-2021} and references therein), but we decided to use an indicator, $\tempor{}$, named \emph{temporality}~\cite{li-natcomms-2020}. Given an ordered sequence of $N_s$ snapshots, $G(t)$, we can associate one value of $\tempor{}$ to each of its $N_s - 1$ pairs of adjacent snapshots. Such a sequence (\ie{} time series) of values can be thought as a fingerprint of $G(t)$ itself. Hence, we can assume that the properties of the distribution of the values of the temporality series can be used as a proxy to characterize the ordering of $G(t)$. In particular, the average value of the temporality $\avgtempor{}$ -- computed over the sequence of $N_s - 1$ values, -- is enough to grasp the main features of the snapshots' ordination.

Leveraging these assumptions, in this work we characterize the persistence of the interactions of several empirical time-varying networks. In particular, we compare the average value of the temporality with the same quantity computed for a random order of the snapshots. The latter can be computed both analytically from the raw data, as well as numerically via sampling the configuration space. We also estimate the boundaries of the configuration space by identifying the configurations corresponding to the maximum and minimum average temporality. Finally, we perform a coarse-graining of the sequences to study the effects of time resolution on the interactions' persistence. 

Our results show that same values of average empirical temporality can stem from different mechanisms of persistence, and that only the comparison with a null model allows to discriminate them. Moreover, we have observed that analysing the system at different (time) resolutions highlights stark differences in the evolution of the persistence at different time scales. Such differences exist even among systems of the same kind, casting doubts on the generally accepted idea that systems of the same type are similar (from our perspective, at least).

\section*{Results}
\label{sec:results}

We divide the characterization of the persistence of the interactions in time-varying networks in two main parts, as explained in the Methods. First, we characterize the ``raw'' sequence (\ie{} without aggregation) by comparing its average temporality with its ``randomly shuffled'' counterpart. Then we repeat the comparison but, this time, using the aggregated version of the sequences. A brief description of the datasets used in our study is available in the Section named ``Data'' of the Methods.

\subsection*{Characterization of unaggregated networks}
\label{ssec:raw_data}

%
%
%
\begin{figure}[t!]
\centering
\includegraphics[width=0.83\textwidth]{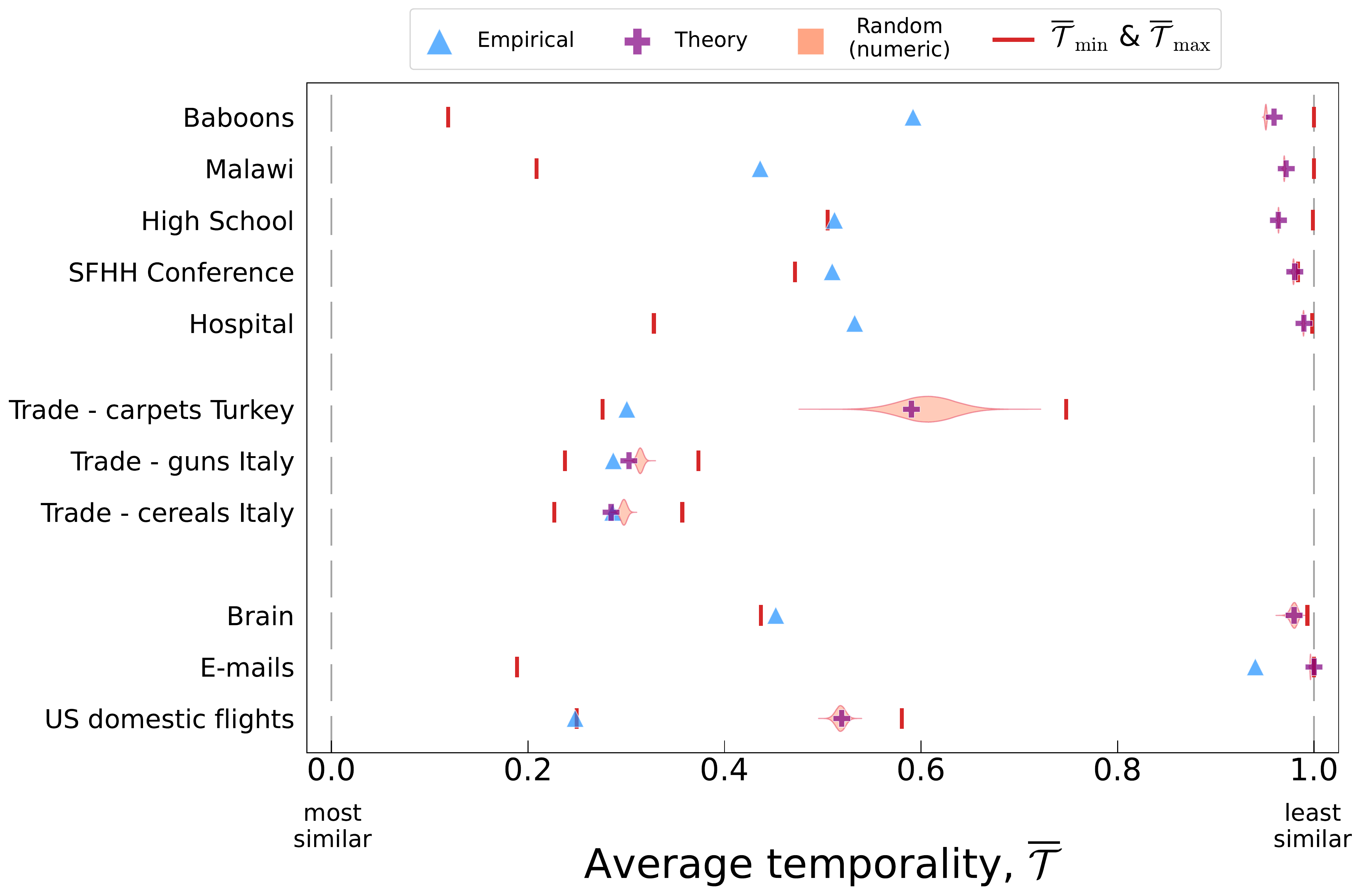}
\caption{Characterization of the average temporality, $\avgtempor$, of the empirical datasets considered in our study. For each dataset, we display the empirical value of $\avgtempor{}$, its theoretical estimation, the distribution (violin plot) of the temporality of randomly shuffled sequences of snapshots, and its maximum and minimum possible values ($\avgtempor_{\max}, \avgtempor_{\min}$). The vertical dashed lines at $\avgtempor = 0.0$ and $\avgtempor = 1.0$ highlight the temporality's theoretical boundaries. To compute the violin plots, we generate $5\times10^6$ randomly shuffled sequences.}
\label{fig:temporality_all_data}
\end{figure}

Figure~\ref{fig:temporality_all_data} summarises the relationships existing between the values of temporality (see Methods) of all the networks considered in our study. More specifically, for each network we report the average value of the temporality computed: %
\begin{inparaenum}[i)]
    \item numerically from the data ($\avgtempor{}$ -- Empirical);
    \item analytically from the data via Eq.~\eqref{eq:average_temp_theo} ($\avgtemportheo{}$ -- Theory);
    \item sampling numerically sequences extracted from the configuration space ($\avgtemporrand{}$, -- Random) (in the text we use $\avgtemporrand{}$ to refer to the median of the distribution);
    \item extracting the sequences corresponding to the maximum and minimum values ($\avgtempor{_{\min}}$,$\avgtempor{_{\max}}$) using the optimization algorithm described in the Methods.
\end{inparaenum}

First of all, we want to remark that for almost all the networks considered in this study, the theoretical estimation $\avgtemportheo{}$ is in good agreement with $\avgtemporrand{}$. Such an agreement means that the shuffling process underlying the computation $\avgtemporrand{}$ fulfils the hypothesis used to derive $\avgtemportheo{}$ and that, more in general, we can use $\avgtemportheo{}$ as a proxy for the estimation of $\avgtemporrand{}$.

Also, Fig.~\ref{fig:temporality_all_data} shows that most of the values of the empiric $\avgtempor{}$ fall within the $[\,0.2, 0.6\,]$ interval, and that empiric sequences have more persistent interactions than the corresponding theoretical estimation (which also represents the randomized counterparts) \ie{} $\avgtempor{} \leq \avgtemportheo{}$. The relative position of $\avgtempor{}$ and $\avgtemportheo{}$, $d_{\text{th}} = \left\lvert \avgtempor{} - \avgtemportheo{} \right\rvert$, provides us with a valuable information. Such a quantity, in fact, determines whether the persistence, if any, stems either from the existence of a sort of permanent set/core of interactions (\ie{} $d_{\text{th}} \sim 0$) or, alternatively, from the existence of short range correlations between temporal-adjacent snapshots which -- in turn, -- controls the lifespan of the interactions (\ie{} $d_{\text{th}} > 0$). Eyeballing at the diagram, we observe that in all the networks (except for the Italian trade ones) $\avgtempor{} \ll \avgtemportheo{}$, implying that the origin of the persistence in the interactions between adjacent snapshots is due to the existence of short-range correlations between temporal-adjacent snapshots.

Another useful indicator is the distance $d_{\min} = \left\lvert \avgtempor{} - \avgtempor{_{\min}} \right\rvert$. We observe that for the networks of \highschooldata{}, \sfhhconfdata{}, \tradturcarpdata{}, \braindata{}, and \usflightsdata{} $d_{\min} \simeq \varepsilon$ with $\varepsilon \rightarrow 0$. Such a trend hints at the existence of some kind of intrinsic optimization behind these networks' organization. For \emaildata{}, instead, the situation is quite the opposite with a system that is almost the least persistent possible. For the \baboonsdata{}, \malawidata{}, \hospitaldata{}, and Italian trade networks we observe bigger values of $d_{\min}$, implying that although the similarities between adjacent snapshots are non negligible (compared to $\avgtemporrand{}$), they are stronger with snapshots distant in time (without having information on how far, though).

Finally, we observe that -- for most networks, -- the theoretical estimation is close to the $\avgtempor{_{\max}}$. Such a proximity implies that each snapshot is as different as possible with the majority of the other snapshots, except for those which are adjacent in the original sequence. In the three trade networks and that of \usflightsdata{} the theoretical estimation lays more far away from the maximum. One possible explanation of such a feature could be related with the fact that these networks have the smallest number of snapshots, $N_s$ (see Table~\ref{tab:datasets}).

Apart from the characterisation of empirical networks, we have computed these metrics for synthetic time-varying networks for which the edges' persistence and the origin of such a persistence can be tuned. In Supplementary Note S3 of the SM, we present different methods for the generation of synthetic datasets in which the persistence of edges is due to the existence of a stable (time-invariant) core of interactions or, alternatively, to the presence of short-range correlations between temporal-adjacent snapshots. The results of our characterisation (see Supplementary Note S3.2 of the SM) provide some insights that argue for the role of different types of persistence on the values of temporality observed in our empirical networks.

\subsection*{Effects of changing the time resolution}
\label{ssec:Agg_wind}

After characterizing the persistence of the interactions, and estimating how special its value is compared with some null hypothesis, we study what are the effects of changing the time resolution on the phenomenology observed. For this reason, we perform a coarse-graining (\ie{} aggregation as described in the Methods) and compute $\avgtempor{}$ as a function of the size of the aggregation window, $\timewind$. One of the goals of studying the aggregation process is to estimate the level of coarse-graining (\ie{} the size of the aggregation window) at which the empirical sequence's order is statistically equivalent to the random one. Said in other terms, we seek to find the point at which the correlations between adjacent snapshots do not play any role for the persistence of the interactions.

%
%
%
\begin{figure}[!ht]
\centering
\includegraphics[width=0.85\textwidth]{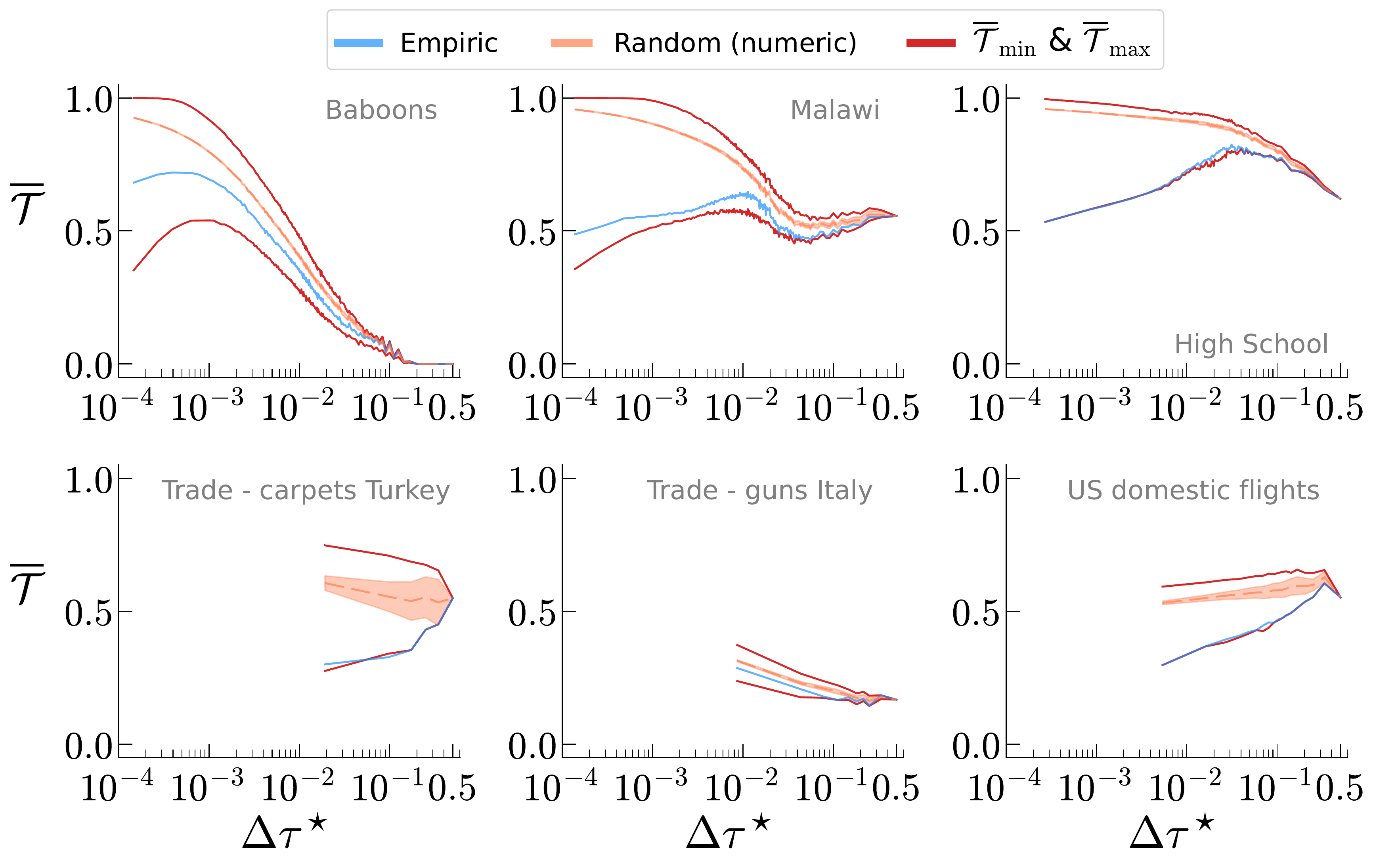}
\caption{Effects of the coarse graining (\ie{} time resolution) on the temporality. We display the values of the empiric temporality, $\avgtempor{}$, of the mean and standard deviation of the distribution of temporality of randomly shuffled sequences, $\avgtempor{_{\text{rand}}}$, and of the maximum and minimum values ($\avgtempor{_{\max}}$, $\avgtempor{_{\min}}$) as a function of the rescaled time resolution $\timewindstar$. Each panel refers to a distinct dataset.}
\label{fig:temporality_vs_aggregation_sel_datasets}
\end{figure}
%

%

%
Figure~\ref{fig:temporality_vs_aggregation_sel_datasets} portrays the behaviour of $\avgtempor{}$ as a function of the aggregation's level, $\timewind$, for six networks whose behaviours encompass the whole spectrum of the phenomenology observed (see Supplementary Fig.~S3 in Supplementary Note S5 for the same picture displaying the whole set of datasets considered in our study). For each network, we display the value of $\avgtempor{}$ of the empiric sequence, the mean (and standard deviation) of the random sampling of the configuration space, and the maximum and minimum values. We decided to not display the theoretical estimation too, since its estimation's accuracy depends on $N_s$ which, in turn, decreases with the coarse-graining (at the coarsest level $N_s = 2$). Finally, to enable the comparison between distinct networks at the same aggregation level, we have to replace $\timewind$ with its rescaled counterpart $\timewindstar =\tfrac{\timewind}{N_s} = \tfrac{n \, t}{N_s}$.

Overall, we observe the following behaviours: %
\begin{inparadesc}
    \item[\baboonsdata{}:] As we aggregate together the snapshots, we observe a rise and fall of the empiric $\avgtempor{}$ and $\avgtempor{_{\min}}$, with a maximum located around $\timewindstar \sim 0.001$ and corresponding to the resolution at which the snapshots are the least similar to each other. As for the non-aggregated data, the empiric temporality is closer to the minimum than the maximum with the random displaying the opposite feature, instead. The relative distance between the random/empiric temporality and the extreme values decreases as we aggregate more, with the four curves merging together around $\timewindstar \sim 0.15$. After that, $\avgtempor{} = 0$ as the network becomes complete/fully connected.
%
%
%
%
    \item[\malawidata{}:] Aggregating the snapshots produces oscillations on the empiric $\avgtempor{}$ with the global maximum located around $\timewindstar \sim 0.015$. Moreover, we notice that the empiric and random values of $\avgtempor{}$ remain always close to $\avgtempor{_{\min}}$ and $\avgtempor{_{\max}}$, respectively. The behaviour of the empiric and random $\avgtempor{}$ suggest that long and short range correlations between snapshots are nearly the same. It is worth mentioning that also the \emaildata{} network displays a similar behaviour [see Supplementary Fig.~S3 of the Supplementary Materials (SM)].
%
%
%
    \item[\highschooldata{}:] We observe an almost perfect overlap between the empiric temporality and the minimum one, and between the random and the maximum one across the whole $\timewindstar$'s range. As for \baboonsdata{}, the empiric temporality has a maximum for $\timewindstar \sim 0.02$ and the four curves merge together around $\timewindstar \sim 0.3$. The \sfhhconfdata{}, \hospitaldata{}, and \braindata{} data display a similar behaviour (see Supplementary Fig.~S3 of the SM).
    \item[\tradturcarpdata{}:] As for the previous case, the empiric and minimum temporality curves overlap almost perfectly across the whole $\timewindstar$'s range. However, the mean value of $\avgtempor{_\text{rand}}$ lays more or less in the middle between $\avgtempor{_{\max}}$ and $\avgtempor{_{\min}}$. Also, the value of $\avgtempor{_{\text{rand}}}$ it not affected too much by the temporal resolution at which we study the system. Finally, the curves overlap with each other only at $\timewindstar \simeq 0.5$.
    \item[\traditagunsdata{}:] We observe a monotonous decrease of $\avgtempor{}$ for all indicators. Moreover, the empiric and mean value of $\avgtempor{_{\text{rand}}}$ are very similar for almost the whole range of $\timewindstar$ values, implying that the phenomenology observed for unaggregated data is not affected by the coarse graining.
    \item[\usflightsdata{}:] We observe a behaviour similar to the Turkish trade dataset one, albeit in this case the temporality is monotonically increasing with $\timewindstar$. Such a behaviour denotes that the snapshots get less similar as we aggregate them.
\end{inparadesc}
Furthermore, it is worth mentioning that for every dataset all the temporality values must coincide for $\timewindstar = 0.5$ (\ie{} when the sequence is made only by two snapshots). Such a phenomenon, stems from the fact that there are only two possible snapshots' orders which correspond to the same temporality.
%
%
%
%
Apart from that, one might expect that the value of $\avgtempor{}$ becomes smaller for higher values of $\timewindstar$ as the snapshot networks becomes more akin to complete networks. However, this is not the case and $\avgtempor{}$ at $\timewindstar = 0.5$ falls within the whole range of possible values with $\avgtempor{} \simeq 1$ for \sfhhconfdata{} and $\avgtempor{} = 0$ for \baboonsdata{}, instead (see Supplementary Fig.~S3 of the SM).

Finally, one feature observed in all datasets is that the distance between $\avgtempor{}$ and $\avgtempor{_{\text{rand}}}$ decreases with $\timewindstar$ and, eventually, goes to zero for $\timewindstar = 0.5$. This phenomenon indicates, according to the analysis performed on the characterisation of non-aggregated sequences, that as the aggregation window increases, a core of fixed interactions emerges.

\section*{Discussion and Conclusion}
\label{sec:conclusion}

The static network paradigm is short to fully mimic the rich phenomenology displayed by dynamics taking place in those systems whose interactions evolve in time. Using a time-varying network paradigm allows to overcome such limitations and, in turns, to attain a better description of the interplay between dynamics and the persistence of evolving interactions~\cite{masuda-prl-2013,masuda-scirep-2016}.

In this work, we used a metric to gauge the interactions' persistence named temporality~\cite{li-natcomms-2020}, and proposed an approach based on statistical physics to characterise the features of several empirical time-varying networks. By comparing the values of the average temporality, $\avgtempor{}$, with its counterpart obtained from a random sampling of the configuration space, we assess how ``special'' the empirical order of snapshots is. Remarkably, we have found that the empiric interactions tend to be more persistent than in a random sequence of snapshots, and that some systems follow some kind of optimisation principle behind the arrangement of their snapshots. Besides, we have studied also the effects of time-resolution -- \ie{} coarse-graining -- $\timewind$ on the temporality and the evolution of the hierarchy between its values computed for empiric, random, and limit ($\avgtempor{_{\max}}, \avgtempor{_{\min}}$) sequences. We have observed how systems belonging to the same category (\eg{} face-to-face interactions) display distinct behaviours as we reduce the time resolution. According to the formalism of statistical physics, such an approach could be extended from the canonical formulation -- in which the temporality (akin to the energy) is not fixed, but the number of pairs of snapshots (a proxy for the number of particles) is, -- to the grand canonical formulation; thus allowing for the characterisation of the datasets with respect to a configurations' space in which the number of snapshots, $N_s$, varies~\cite{cimini-nat_rev_phys-2019}.

The comparison of the empiric value of $\avgtempor{}$ and the mean of the distribution of temporality obtained from the random sampling of the configuration space, $\avgtemporrand{}$, provides us with a valuable information about the origin of the interactions' persistence. In particular, we can ask ourselves whether the latter is due either to the existence of intrinsic (temporal) correlations between adjacent snapshots (\ie{} memory)~\cite{williams-natcomms-2022}, or to the presence of a set of persistent interactions whose existence is not affected by the temporal order of the snapshots.
Two examples of these extreme configurations are the Italian trade datasets and the \braindata{} one. Both exhibit a low value of $\avgtempor{}$ implying the persistence of the interactions between adjacent snapshots. However, trade datasets display also low values of $\avgtemporrand{}$, whereas \braindata{} has $\avgtemporrand{} \sim 1$. Such a difference highlights the existence of a stable, persistent, set of interactions for the former networks, whereas in the latter network short-range memory prevails.

Finally, the methodology presented in this work has some technical -- and conceptual, -- limitations which could become the subject of further studies.
For instance, one could generalise the estimation of $\avgtempor{}$ to account for the existence of correlations in the probability that the same edge exists in adjacent snapshots (which is one of the hallmarks of real time-varying networks)~\cite{book-masuda-2020,williams-natcomms-2022,ferguson-epjds-2022}. Another possibility is to leverage the information carried by the set of each link's \emph{local} values of temporality. Among the potential extensions, one is to explore the interplay between the edges belonging to the time-varying backbone~\cite{gemmetto-arxiv-2017,kobayashi-natcomm-2019} and those contributing to the temporality observed in empirical systems. Studying the evolution of topological descriptors could help to grasp the behaviour of the temporality when one changes the size of the aggregation window (\eg{} to find a ``characteristic'' time-scale). Finally, the approach presented in this work could be used as a basis for the design of a method (akin to a configuration model~\cite{bender-j_comb_theo-1978,fosdick-siam_rev-2018}) to generate time-varying networks with a given value of average temporality, and use it to test the role of persistent interactions on dynamical processes in a more controlled way.

\section*{Methods}
\label{sec:methods}

A time-varying network $G(\mathcal{N},\mathcal{E}(t))$ with $N \equiv \modu{\mathcal{N}}$ vertices (or nodes) is defined as a set of interaction's triples $e \equiv (i,j,t) \in \mathcal{E}(t)$ where $i, j \in \{1, \ldots, N \}$ are the indices of the interacting vertices, and $t$ denotes the time at which such an interaction occurs (we are assuming that the set of vertices, $\mathcal{N}$, does not change over time). Time-varying networks can be thought also as a sequence of $N_s$ graphs (snapshots), $G(t) \equiv G(\mathcal{N},\mathcal{E}_t)$, each made only by those interactions occurring at the same, discrete, timestep $t$~\cite{book-masuda-2020,holme-phys_rep-2012}. In this section, first we introduce the concept of \emph{temporality} $\tempor$ and how to estimate its average value, $\avgtempor$, over the sequence $G(t)$. Then, we describe how we merge together adjacent snapshots, shuffle the elements of $G(t)$, as well as find the sequences $G^{\min}(t)$ and $G^{\max}(t)$ corresponding to the maximum and minimum values of $\avgtempor$ within the configuration space.

\subsection*{Temporality}
\label{ssec:temporality}

Given a pair of snapshots graphs $G_m, G_n \in G(t)$ (with $m,n \in \{1, \ldots, N_s \}$), we define its \emph{temporality}, $\tempor_{m,n}$, as:
\begin{equation}
\label{eq:temporality_def}
\tempor_{m,n} = \frac{\sum\limits_{i,j = 1}^{N} \modu{ a_{i,j}(m) - a_{i,j}(n)}}{\sum\limits_{i,j} \max \left\lbrace a_{i,j}(m) , a_{i,j}(n) \right\rbrace } \,,
\end{equation}
where $a_{i,j}(m)$ is an element of the adjacency matrix $\mathcal{A}_m$ of the graph $G_m$~\cite{latora-book-2017}. In particular, $a_{i,j}(m)$ is equal to one if vertices $i$ and $j$ are connected in graph $G_m$, and is equal to zero otherwise. Equation~\eqref{eq:temporality_def} quantifies nothing else than the ratio between the number of distinct edges of $G_m$ and $G_n$ (\ie{} those existing in one snapshot but not in the other), divided by the number of common and non-common edges. We can rewrite Eq.~\eqref{eq:temporality_def} in terms of the set of edges $\mathcal{E}$ as:
\begin{equation} 
\label{eq:temporality}
\tempor_{m,n} = \frac{\left\lvert\bigcup_{m,n}\right\rvert -\left\lvert\bigcap_{m,n}\right\rvert}{\left\lvert\bigcup_{m,n}\right\rvert} =  1 - \frac{\left\lvert\bigcap_{m,n}\right\rvert}{\left\lvert\bigcup_{m,n}\right\rvert} \,,
\end{equation}
where $\left\lvert\bigcup_{m,n}\right\rvert$ is the size of the union of the edges' sets of graphs $G_m$ and $G_n$ (\ie{} $\bigcup_{m,n} \equiv \mathcal{E}_m \cup \mathcal{E}_n$), whereas $\left\lvert\bigcap_{m,n}\right\rvert$ is the size of the intersection of those sets, (\ie{} $\bigcap_{m,n} \equiv \mathcal{E}_m \cap \mathcal{E}_n$).

We can use the definition of temporality to gauge the volatility of the interactions of a time-varying network. To this aim, we define the average value of the temporality, $\avgtempor$, for the whole time-varying network, $G(t)$ (\ie{} the whole set of snapshots), as follows:
\begin{equation} 
\label{eq:average_temp}
\avgtempor = \frac{1}{N_s - 1} \sum_{m=1}^{N_s-1} 1 - \frac{\left\lvert\bigcap_{m,m+1}\right\rvert}{\left\lvert\bigcup_{m,m+1}\right\rvert} = 1 - \frac{1}{N_s-1} \sum_{m=1}^{N_s-1} \frac{\left\lvert\bigcap_{m,m+1}\right\rvert}{\left\lvert\bigcup_{m,m+1}\right\rvert}\,,
\end{equation}
where $m, m+1 \in \{1, \ldots, N_s - 1 \}$ denote the indices of two temporal-adjacent snapshots of the time-varying network $G(t)$. According to the above definition, the temporality values span from $0$ for networks having always the same edges, to $1$ for completely different networks.

It is possible to estimate analytically $\avgtempor$ by assuming that %
\begin{inparaenum}[i)]
    \item the microscopic process governing the existence of an interaction between two vertices at time $t$ is independent on both the existence of other interactions at the same time, as well as on the occurrence of such an interaction in the past.
    \item The microscopic process governing the existence of an interaction does not change over time.
    \item The network has a constant size (\ie{} $N$ does not vary over time).
\end{inparaenum}

Following these assumptions, the average number of interactions (edges) in a snapshot of $G(t)$, $\avg{\mathcal{L}}$, can be expressed as:
\begin{equation}
\label{eq:avg_nr_edges_theo}
\avg{\mathcal{L}} = \frac{N \, (N-1)}{2} \, \avg{x_{ij}} \,,
\end{equation}
where $x_{ij}$ is the probability that an edge between nodes $i$ and $j$ exists in any of the snapshots, and $\avg{x_{ij}}$ is the average of such a probability over all the $\tfrac{N(N-1)}{2}$ possible edges. By leveraging the independence of interactions taking place at different snapshots, we can express the average number of edges belonging to two time adjacent snapshots (\ie{} to the intersection of their edges' sets), $\bigl\langle \left\lvert \cap \right\rvert \bigr\rangle$, as:
\begin{equation}
\label{eq:avg_nr_inter_edges}
\bigl\langle \left\lvert \cap \right\rvert \bigr\rangle = \frac{N \, (N-1)}{2} \, \avg{x_{ij \, \in \, \cap}} = \frac{N \, (N-1)}{2} \, \avg{x_{ij}^2} \,,
\end{equation}
where $x_{ij \, \in \, \cap}$ is the probability that the edge $e(i,j)$ belongs to the intersection of the edges' sets of all the snapshots which, according to the independence of the probabilities postulated above, is equal to the product of the individual probabilities $x_{ij}$.

Finally, as the microscopic process governing the existence of the interactions does not change over time (\ie{} it is independent of $t$), we can rewrite Eq.~\eqref{eq:average_temp} in terms of Eqs.~\eqref{eq:avg_nr_edges_theo} and \eqref{eq:avg_nr_inter_edges}, yielding:

\begin{equation}
\label{eq:average_temp_theo}
\avgtemportheo{} = 1 - \frac{1}{N_s - 1} \sum_{m=1}^{N_s - 1} \frac{\left\lvert\bigcap_{m,m+1}\right\rvert}{\left\lvert\bigcup_{m,m+1}\right\rvert} = 1 - \left\langle \frac{\left\lvert \cap \right\rvert}{\left\lvert \cup \right\rvert} \right\rangle = 1 - \left\langle \frac{\left\lvert \cap \right\rvert }{2 \, \mathcal{L} -  \left\lvert \cap \right\rvert } \right\rangle \approx 1 - \frac{\bigl\langle \left\lvert \cap \right\rvert \bigr\rangle}{2 \, \avg{\mathcal{L}} - \bigl\langle \left\lvert \cap \right\rvert \bigr\rangle} = 1 - \frac{\bigl\langle x_{ij}^2\bigr\rangle}{2 \, \avg{x_{ij}} - \bigl\langle x_{ij}^2 \bigr\rangle} \,.
\end{equation}

%
%
%
Hence, the theoretical estimation of $\avgtempor$ can be written just in terms of $x_{ij}$. However, we want to stress that the approximation is valid only if both the variance of $\mathcal{L}$ and $\left\lvert \cap \right\rvert$, as well as the covariance between the same quantities (and its higher order moments), are both negligible. For more details, see~Supplementary Note S1 of the SM. Hence, to estimate $\avgtempor$ of an empirical network we just have to estimate the probability of appearance of each possible edge. Such a probability can be computed directly from the data as:
\begin{equation}
\label{eq:empiric_edge_prob}
x_{ij} = \frac{N_s(i,j)}{N_s} \,,
\end{equation}
%
%
%
%
where $N_s(i,j)$ is the number of snapshots in which the edge $e(i,j)$ exists. Ensuring a correct interpretation of these results entails some caveats. For instance, the assumption of the interactions' statistical independence between snapshots is, in general, not true for empirical systems. Such an issue exists when the system possesses some sort of memory (even just at short range)~\cite{williams-natcomms-2022}. Nonetheless, as we will see, for uncorrelated sequences our theoretical framework estimates quite well the values of $\avgtempor$. One potential extension of our framework (see~Supplementary Note S2 of the SM) is to compute $x_{ij}$ using the so-called \emph{activity driven model}: a popular model used to generate time-varying networks whose structure resembles those of the empirical ones~\cite{perra-scirep-2012}.

\subsection*{Aggregation, shuffling, and finding extreme values of temporality}
\label{ssec:aggreg_shuf_maxmin}

In the following, we describe how given a sequence of snapshots, $G(t)$, we aggregate, shuffle, and re-order it. Before describing these processes, it is worth mentioning that we perform some pre-processing on the empirical sequences. Specifically, for each dataset we remove those snapshots corresponding to empty graphs (\eg{} night recordings of face-to-face interactions). Such operation is justified by the fact that empty graphs do not contribute to the temporality. Pruning inactivity periods from the data leads to a re-definition of the concept of time itself, converting the variable $t$ from a time into a descriptor of the position of the snapshot along the sequence.

\subsubsection*{Aggregation}

\begin{figure}[t!]
\centering
\includegraphics[angle=-90,width=0.9\textwidth]{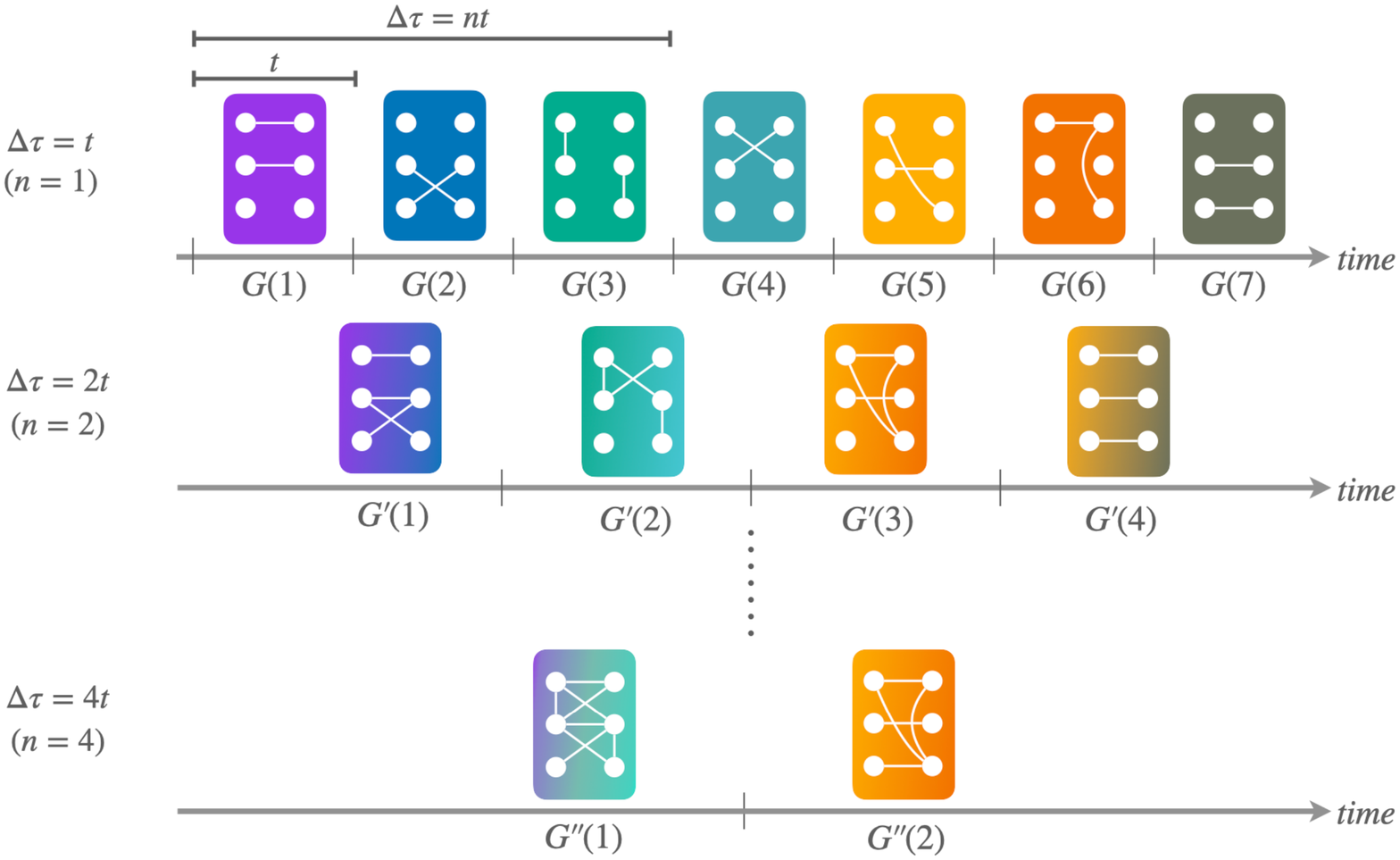}
\caption{Schematic representation of the aggregation process. Given a sequence of snapshots, $G(t)$, we aggregate it by merging together groups of snapshots of size $\timewind = nt$ with $n \in \mathbb{N}$. Each row accounts for a different level of aggregation: $n=1$ (top), $n=2$ (middle), and $n=4$ (bottom).}
\label{fig:aggregacion_temporal}
\end{figure}

One important aspect of the characterization of time-varying networks is the \emph{resolution} at which we study them~\cite{tang-pre-2010}. Given a sequence of snapshots $G(t)$, we can convert it into a new sequence $G^\prime (\tilde{t})$ by aggregating its elements into groups of size $\timewind = nt$ with $n \in \mathbb{N}$ (\ie{} we create its time-wise coarse graining). There are several ways of generating $G^\prime(\tilde{t})$~\cite{holme-phys_rep-2012}. Here, we consider an aggregation approach analogous to that used by Tang \etal{}~\cite{tang-pre-2010} consisting in creating a projection network corresponding to the union of the edges' sets of the snapshots' group of size $\timewind$ (see Fig.~\ref{fig:aggregacion_temporal}).

According to such a projection method, given a sequence of snapshots $G(t)$ and a temporal resolution $\timewind$, the resulting aggregated sequence $G^\prime (\tilde{t})$ will be made of  $\tfrac{N_s}{\timewind}$ snapshots (chunks). If $N_s$ is not an integer multiple of $\timewind$, we apply periodic boundary conditions to the sequence and ``complete'' the last chunk of snapshots by adding those located at the beginning of the sequence (see middle row of Fig.~\ref{fig:aggregacion_temporal}). That said, it is worth stressing that distinct values of $n$ (or, equivalently, $\timewind$) may lead to aggregated sequences with the same number of snapshots (chunks). In such a case, we keep only the values of $n$ minimizing the number of snapshots needed to complete the last chunk.
%
%
%

\subsubsection*{Shuffling}

One way to get rid of the correlations of a time series is to shuffle its elements. Similarly, we can get rid of the correlations existing between adjacent snapshots of the sequence $G(t)$ by randomly shuffling their positions, giving rise to a new sequence $G^\prime(t)$. The snapshots' random arrangement is nothing else than the generation process laying behind our configuration space, as explained in the Introduction. Given a sequence of graphs $G(t)$ with $N_s$ snapshots, it is -- in general -- computationally unfeasible to estimate exactly descriptors like the average or the standard deviation of an indicator (\eg{} of $\tempor{}$) via exploring the whole configuration space, as its size scales with $N_s!$. For this reason, we compute these descriptors by sampling the configuration space generating a computationally-feasible number of realizations ($5\times10^6$) of the random reshuffling process.

Finally, when studying the effects of time resolution (\ie{} aggregation), we aggregate the sequence first and perform the reshuffling then, thus exploring the configuration space of the aggregated sequences.

\subsubsection*{Optimization algorithm}

In the section entitled Temporality, we have mentioned that $\avgtempor{} \in [0,1]$. However, such boundaries are just theoretical and correspond to very peculiar configurations. To gauge the effective boundaries of the temporality, we need to identify the arrangements (\ie{} sequences) corresponding the maximum and minimum values of $\avgtempor{}$. This optimization problem shares many traits with the so-called \emph{Travelling Salesman Problem}~\cite{JUNGER1995225}. Specifically, each of the $N_s$ snapshots of the sequence $G(t)$ can be thought as the vertex of a graph, and the temporality computed between each of the $\tfrac{N_s\,(N_s - 1)}{2} $ snapshots' pairs can be thought as the distance between vertices. Therefore, the aforementioned optimization problem consists in finding the open chain (\ie{} snapshots' sequence) maximizing -- or minimizing, -- its total length (a proxy for the average temporality). However, the solution of such an optimization problem is NP hard. To overcome such a limitation, we use a heuristic algorithm based on the \emph{Kruskal's algorithm} (used to extract the minimum spanning tree of a graph)~\cite{cormen-book-2001}, allowing us to obtain an approximate solution for our problem.

Given the set of snapshots, $\mathbb{G} \equiv \{ G_i \}$ with $G_i \in G(t) \; \forall \, i = \{1, \ldots, N_s \}$, our heuristic algorithm works as follows:
\begin{enumerate}
    \item Compute using Eq.~\eqref{eq:temporality} the elements of the set $\mathbb{T} \equiv \!\bigl\lbrace \tempor_{m,n} \bigr\rbrace_{m,n = 1}^{N_s} \, \forall \, m,n \in \{1, \ldots, N_s \}, \, m \neq n$; and store its values. Then, sort the elements of set $\mathbb{T}$ in ascending (or descending) order.
    \item For each $T_{m,n} \in \mathbb{T}$, we add the corresponding edge $\left(G_m, G_n \right)$ if and only if $G_m$ and $G_n$ are either both isolated vertices (graphs), or members of distinct chains and each vertex is connected with at most another vertex.
    \item Repeat the above operation until the system is made by a single chain with $N_s - 1$ edges.
\end{enumerate}
Inverting the sorting order allows to determine either the snapshots' sequence $G^{\min}(t)$ corresponding to $\avgtempor_{\min}$, or the sequence $G^{\max}(t)$ corresponding to $\avgtempor_{\max}$. As mentioned previously, the non-Markovian nature of the algorithm leaves room to the chance that different choices, locally not optimal, could lead to a better final outcome (\ie{} the solution found is a local optimum, but not necessarily a global one). One possible workaround is to add the edge $(G_m, G_n)$ with a probability computed using a simulated annealing technique \cite{newman-book_compphys-2012}. Finally, we would like to mention that the above algorithm could be modified to reduce the computation time.

\subsection*{Data}
\label{ssec:data}

We consider eleven time-varying networks grouped in the following categories:
\begin{description}
    \item[Face-to-face] Five networks obtained from the Sociopatterns' repository~\cite{sociopatterns} Specifically:%
    \begin{description}
        \item[\baboonsdata{}] The interactions occurring among a group of Guinea baboons living in an enclosure of a Primate Center in France recorded between June 13\textsuperscript{th} and July 10\textsuperscript{th} of 2019~\cite{gelardi-proc_roy_soc_a-2020}. Note: the great number of snapshots ($\sim 40,000 $) of the dataset make unfeasible the sampling and the optimization over the configuration space. For this reason, we need to make a 2-snapshots aggregation even for the unaggregated characterization.
        \item[\malawidata{}] Observational contact data collected for 86 individuals living in a village in rural Malawi~\cite{ozella-epj_ds-2021}. Note: the great number of snapshots ($\sim 40,000 $) of the datasets make infeasible the sampling and the optimization over the configuration space. For this reason, we need to make a 2-snapshots aggregation even for the unaggregated characterization.
        \item[\highschooldata{}] Contacts and friendship relations between students in a high school in Marseilles (France) recorded during December 2013~\cite{fournet-pone-2014}.
        \item[\sfhhconfdata{}] Interactions among the participants to the \textsc{SFHH} conference in Nice (France) which took place between June 4\textsuperscript{th} and 5\textsuperscript{th} of 2009~\cite{genois-epj_ds-2018}.
        \item[\hospitaldata{}] Contacts between patients, patients and health-care workers (HCWs), and between HCWs in a hospital ward in Lyon (France) recorded between December 6\textsuperscript{th} and December 10\textsuperscript{th} of 2010~\cite{vanhems-pone-2013}.
    \end{description}
    %
    %
%
    %
    %
    \item[Trade] Three star networks describing the export relationships over a specific good (commodity) occurring between one country and all the other countries in the world, extracted from the UN-COMMTRADE database~\cite{uncommtrade} (see Supplementary Note S4 of the SM for the details). Specifically:%
    \begin{description}
        \item[\tradturcarpdata{}] Exports of ``\emph{carpets, carpeting and rugs, knotted}'' taking place between years 1962 and 2020.
        \item[\traditagunsdata{}] Exports of ``\emph{Arms and ammunition; parts and accessories thereof}'' taking place between January 2010 and December 2020.
        \item[\traditaceradata{}] Exports of ``\emph{cereals}'' taking place between January 2010 and December 2020.
    \end{description}
    \item[Other] Three networks of various types. Specifically:%
    \begin{description}
        \item[\braindata{}] The functional brain network extracted from the EEG $\beta$ band activity recorded in several Regions of Interest (ROIs) during a motor task~\cite{de_vico_fallani-j_phys_a-2008}.
        \item[\emaildata{}] The network of e-mail exchanges between members of the US Democratic Party during the $2016$ Democratic National Committee~\cite{Emails1,Emails2}.
        %
        %
%
        %
        %
        \item[\usflightsdata{}] The network of domestic flights operated within the USA taking place between January 1990 and December 2021 (see Supplementary Note S4 of the SM for the details).
    \end{description}
\end{description}
Table~\ref{tab:datasets} presents a summary of the main properties of the above networks.

%
%
\begin{table}[!ht]
\caption{Main characteristics of the datasets considered. For each network, we report the number of nodes, $N$, the number of snapshots $N_s$, the temporal resolution, $\Delta t$, the number of distinct interactions in the aggregated network, $\ktot$, the average temporality, $\avgtempor{}$, and the edges' density average over the snapshots' set, $\avg{\rho}$. Finally, we report the data's bibliographic source.}
\label{tab:datasets}
\centering
\resizebox{0.8\linewidth}{!}{
\begin{tabular}{rrrrrrrc}
\hline
Dataset & \centering $N$ & \centering $N_s$ & \centering $\Delta t$ & \centering $\ktot$ & \centering $\avgtempor{}$ & \centering $\avg{\rho}(\times 10^{-4})$ & Source \\\hline
 & \multicolumn{7}{c}{\textsc{face-to-face}} \\\cline{2-8}
\baboonsdata{} & 13 & 40,845$\star$ & \multirow{5}{*}{20 s} & 78 & 0.592 & 287.47 & \cite{gelardi-proc_roy_soc_a-2020}\\
\malawidata{} & 86 & 43,437$\star$ &  & 347 & 0.436 & 8.51 & \cite{ozella-epj_ds-2021}\\
\highschooldata{} & 327 & 7,374 &  & 5,818 & 0.512 & 4.80 & \cite{fournet-pone-2014}\\
\sfhhconfdata{} & 403 & 3,508 &  & 9,565 & 0.510 & 2.47 & \cite{genois-epj_ds-2018}\\
\hospitaldata{} & 75 & 9,452 &  & 1,139 & 0.532 & 12.36 & \cite{vanhems-pone-2013}\\
\cline{2-8}
 & \multicolumn{7}{c}{\textsc{trade}} \\\cline{2-8}
\tradturcarpdata{} & 207 & 52 & 1 year & 206 & 0.301 & 34.29 & \multirow{3}{*}{\cite{datos_trade,uncommtrade}}\\
\traditagunsdata{} & 156 & 116 & \multirow{2}{*}{1 month} & 155 & 0.287 & 59.92 & \\
\traditaceradata{} & 157 & 108 &  & 156 & 0.286 & 60.41 & \\
\cline{2-8}
 & \multicolumn{7}{c}{\textsc{other}} \\\cline{2-8}
\braindata{} & 16 & 396 & $\frac{1}{200}$ s & 120  & 0.452 & 395.99 & \cite{de_vico_fallani-j_phys_a-2008}\\
\emaildata{} & 1,890 & 19,380 & 1 s & 4,383 & 0.940  & 0.01 & ~\cite{Emails1,Emails2}\\
\usflightsdata{} & 1,677 & 371 & 1 month & 25,890 & 0.248 & 24.16 & \cite{datos_usair,bureau_trans_stats} \\
\hline
\end{tabular}
} 
\end{table}
%


%
%
\bibliography{biblio}

\section*{Acknowledgements}

The authors thank N. Masuda for his comments on a preliminary version of the manuscript, N. Perra for his help with the activity driven model, and F. De Vico Fallani for providing the brain data.

AC acknowledges the support of the Spanish Ministerio de Ciencia e Innovaci\'on (MICINN) through Grant IJCI-2017-34300. AC acknowledges the support of the European Research Council (ERC) under the European Union's Horizon 2020 research and innovation programme 
(Grant agreement No. 803860). AC, AA, and FB acknowledge the support of the DEIM Department of the University Rovira i Virgili via the ``Investigador Activo'' funds.
AA acknowledges financial support from Spanish MINECO (Grant No.\ PGC2018-094754-B-C2), from Generalitat de Catalunya (grant No.\ 2017SGR-896 and 2020PANDE00098), Universitat Rovira i Virgili (grant No.\ 2019PFR-URV-B2-41), Generalitat de Catalunya ICREA Academia, and the James S. McDonnell Foundation (grant \#220020325). LMF and JGG acknowledge financial support from the Departamento de Industria e Innovación del Gobierno de Arag\'on y Fondo Social Europeo (FENOL group E36\_20R), and from grant PID2020-113582GB-I00 funded by MCIN/AEI/10.13039/501100011033.
FB acknowledges the support from Departamento de Industria e Innovaci\'on del Gobierno de Arag\'on y Fondo Social Europeo through projects No. E30\_17R (COMPHYS group) and doctoral fellowship, and the financial support from the Spanish Ministerio de Ciencia e Innovaci\'on (MICINN) through grant PGC2018-094684-B-C22.

Numerical analysis has been carried out using the NumPy and NetworkX Python packages~\cite{oliphant-book-2006,vanderwalt-compscieng-2011,hagberg-scipy-2008}. Graphics have been prepared using the Matplotlib Python package~\cite{hunter-matplotlib-2007}.

\section*{Author contributions statement}

AC and AA secured funding; AC, AA, and JGG designed the study; FB and MF performed the modelling; AC contributed with the data; FB and AC performed the analysis; All authors analysed the results; FB and AC wrote the paper; AC and JGG prepared the graphics. All authors read, reviewed, and approved the final manuscript.

\section*{Additional information}

\begin{description}
\item[Competing interests]
The authors declare no competing interests.
\item[Availability of data and materials]
The data and code used to generate the US domestic flights and\newline\commtrade{} networks are available at:\newline \url{https://cardillo.web.bifi.es/data.html#flights} and\newline \url{https://cardillo.web.bifi.es/data.html#trade}. All the other data used in this study are publicly available (see Table~\ref{tab:datasets} for the bibliographic sources).
\end{description}

\end{document}